\documentstyle[prd,aps,floats,tabularx,epsfig]{revtex}
\newcommand{\be}{\begin{equation}}
\newcommand{\ee}{\end{equation}}
\newcommand{\bea}{\begin{eqnarray}}
\newcommand{\eea}{\end{eqnarray}}

\begin{document}          
\setlength{\unitlength}{1mm}
\twocolumn[\hsize\textwidth\columnwidth\hsize\csname@twocolumnfalse\endcsname
\title{Early-universe constraints on Dark Energy}
\author{Rachel Bean$^\sharp$, Steen H. Hansen$^\flat$ 
and Alessandro Melchiorri$^\flat$}
\address{ $^\sharp$ Theoretical Physics, The Blackett Laboratory, Imperial
 College, Prince Consort Road, London, U.K.\\
$^\flat$ NAPL, University of Oxford, Keble road, OX1 3RH, Oxford, UK}
\maketitle

\begin{abstract}
In the past years 'quintessence' models have been considered which can
produce the accelerated expansion in the universe suggested by recent
astronomical observations.  One of the key differences between
quintessence and a cosmological constant is that the energy density in
quintessence, $\Omega_\phi$, could be a significant fraction of the
overall energy even in the early universe, while the cosmological
constant will be dynamically relevant only at late times.  We use
standard Big Bang Nucleosynthesis and the observed abundances of
primordial nuclides to put constraints on $\Omega_\phi$ at
temperatures near $T \sim 1MeV$.  We point out that current
experimental data does not support the presence of such a field,
providing the strong constraint $\Omega_\phi(\mbox{MeV}) < 0.045$ at
$2\sigma$ C.L. and strengthening previous results.  We also consider
the effect a scaling field has on CMB anisotropies using the recent
data from Boomerang and DASI,
providing the CMB constraint $\Omega_\phi \le 0.39$ at $2\sigma$
during the radiation dominated epoch.
\end{abstract}
\bigskip
]

{\it Introduction.} 
Recent astronomical observations \cite{super1} suggest that
the energy density of the universe is dominated by a dark energy
component with negative pressure which causes the expansion rate of
the universe to accelerate. One of the main goals for cosmology, and for
fundamental physics, is ascertaining the nature of the dark energy
\cite{dark}.

In the past years scaling fields have been considered which 
can produce an accelerated expansion in the present epoch.  The scaling
field is known as ``quintessence'' and a vast category of ``tracker''
quintessence models have been created (see for example
\cite{quint,brax} and references therein), in which the field 
approaches an attractor solution at early times, with its
energy density scaling as a fraction of the dominant component.
The desired late time accelerated expansion behaviour is then 
set up independently of initial conditions, with the quintessential
field dominating the energy content. 

Let us remind the reader of the two key differences between the general
 quintessential model and a cosmological constant: firstly, 
for quintessence, the equation-of-state parameter $w_{\phi}=p/\rho$ 
varies in time,  usually approaching a present value $w_0 <-1/3$, 
whilst for the cosmological constant remains fixed at $w_{\Lambda}=-1$.
Secondly, during the attractor regime 
the energy density in quintessence $\Omega_{\phi}$ is, in general, a
significant fraction of the dominant component 
whilst $\Omega_{\Lambda}$ 
is only comparable to it at late times.

Future supernovae luminosity distance data, as might be 
obtained by the proposed SNAP satellite, will probably have 
the potential to discriminate between different dark energy theories 
\cite{jochen}. 
These datasets 
will only be able to probe the late time behaviour of the 
dark energy component at red-shift $z < 2$, however.
Furthermore, since the luminosity distance depends on $w$ through a 
multiple integral relation, it will be difficult to infer a precise 
measurement of $w(z)$ from these datasets alone \cite{maor}.

In this {\it Letter} we take a different approach to the problem,
focusing our attention on the early time
behaviour of the quintessence field, when the tracking regime is
maintained in a wide class of models, and $\Omega_\phi$ is a significant
($\ge0.01$, say) fraction of the overall density.

In particular, we will use standard big bang nucleosynthesis and the
observed abundances of primordial nuclides to put constraints on the
amplitude of $\Omega_{\phi}$ at temperatures near $T \sim 1 $MeV.  The
inclusion of a scaling field increases the expansion rate of the
universe, and changes the ratio of neutrons to protons at freeze-out
and hence the predicted abundances of light elements.

The presence of this field in the radiation dominated regime 
also has important effects on the shape of the
spectrum of the cosmic microwave background anisotropies.
We use the recent anisotropy power spectrum data obtained
by the Boomerang \cite{Boom2} and DASI~\cite{Dasi} 
experiments to obtain further, independent constraints on 
$\Omega_{\phi}$ during the radiation dominated epoch.


There are a wide variety of quintessential models; we
limit our analysis to the most general ones, with attractor solutions
established well before nucleosynthesis. 


More specifically, we study a tracker model based on the exponential
potential $V=V_0e^{-\lambda \phi}$ \cite{attractor}.
If the dominant component scales as $\rho_n=\rho_0({a_0 \over a})^n$,
then the scaling field eventually approaches an attractor solution, and
its fractional energy density is given by $\Omega_{\phi}={n \over
\lambda^2}$.  However, the pure exponential potential, since it simply
mimics the scaling of the dominant matter in the attractor regime,
cannot produce an accelerated expanding universe in the
matter dominated epoch.

\begin{figure}[thb]
\begin{center}
\epsfxsize=6.2cm
\epsfysize=7.2cm
\epsffile{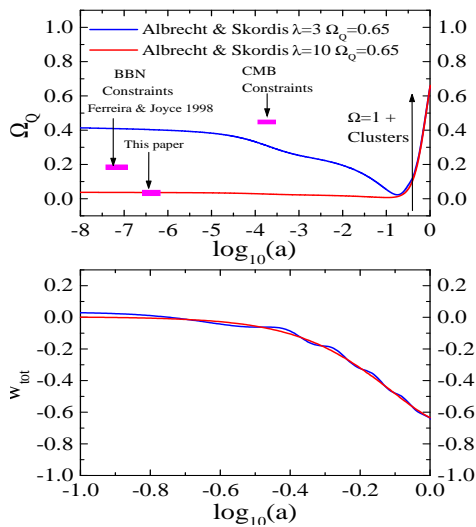}
\end{center}
\caption{Top panel:
Time behaviour of the fractional energy density $\Omega_{\phi}$ 
for the Albrecht and Skordis model together with the constraints
presented in the paper. The parameters of the models are 
(assuming $h=0.65$ and $\Omega_{\phi}=0.65$) 
$\lambda=3$, $\phi_0=87.089$, $A=0.01$ and $\lambda=8$, 
$\phi_0=25.797$, $A=0.01$.
Bottom panel: Time behaviour for the overall equation of state
parameter $w_{tot}$ for the two models. Luminosity distance data
will not be useful in differentiating the two models.}
\label{figomega}
\end{figure}

Therefore, we focus our attention on a recently proposed model by
Albrecht and Skordis (referred to as the AS model from herein) \cite{AS},
motivated by physics in
the low-energy limit of M-theory, which includes a factor
in front of the exponential, so that it takes the form $V=V_0 \left(
(\phi_0-\phi)^2+A \right) e^{-\lambda \phi}$.
The prefactor introduces a small minimum in the potential.
When the potential gets trapped in this minimum its kinetic energy
disappears,
triggering a period of accelerated expansion, which never ends 
if $A \lambda^2 <1$ \cite{BBM}.

In Fig.~\ref{figomega} we introduce and summarise the main results
of the paper. In the figure, the BBN constraints obtained
in section 2, and the CMB constrains obtained in section 3 are shown
together
with two different versions of the AS model which both satisfy
the condition $\Omega_\phi=0.65$ today.

\medskip
{\it Constraints from BBN.}
In the last few years important experimental progress 
has been made in the measurement of light element primordial abundances.
For the $^4He$ mass fraction, $Y_{\mbox{He}}$, two marginally compatible 
measurements have been obtained from regression against zero metallicity
in blue compact galaxies. A low value  $Y_{\mbox{He}} =
0.234 \pm 0.003$ \cite{helow} and a high one $Y_{\mbox{He}} =
0.244 \pm 0.002$ \cite{hehigh} give realistic bounds.
We use the high value in our analysis; if one instead considered
the low value, the bounds obtained would be even stronger.

Observations in different quasar absorption line
systems give a relative abundance for deuterium, critical in fixing the
baryon fraction, of $D/H=(3.0 \pm 0.4) \cdot 10^{-5}$
\cite{delow}.
Recently a new measurement of deuterium in the damped Lyman-$\alpha$
system was presented~\cite{newdeut}, leading to the weighted abundance
$D/H=(2.2 \pm 0.2) \cdot 10^{-5}$. We use the value from~\cite{delow}
in our analysis; the use of~\cite{newdeut} leads to even stronger bound.

\begin{figure}[thb]
\begin{center}
\epsfxsize=6.2cm
\epsfysize=6.2cm
\epsffile{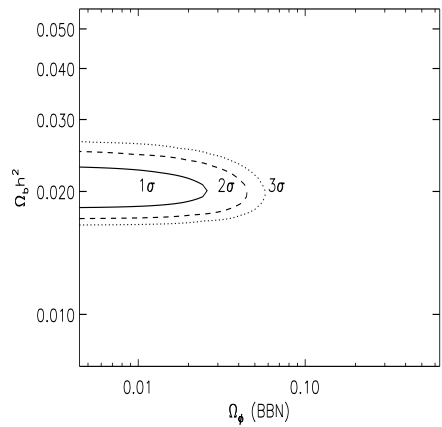}
\end{center}
\caption{$1,2$ and $3 \sigma$ likelihood contours in the $(\Omega_b
h^2, \Omega_\phi (1 \mbox{MeV}))$ parameter space derived from $^4He$
and $D$ abundances.}
\label{figbbn}
\end{figure}

In the standard BBN scenario, the primordial abundances are a function
of the baryon density $\eta \sim \Omega_bh^2$ only. 
To constrain the energy density of a primordial field a $T\sim$ MeV,
we modified the standard BBN code~\cite{kawano}, including the
quintessence energy component $\Omega_\phi$. 
We then performed a likelihood analysis in
the parameter space $(\Omega_bh^2,\Omega_{\phi}^{BBN})$ 
using the observed abundances $Y_{\mbox{He}}$ and $D/H$.
In Fig.~\ref{figbbn} we plot the $1,2$ and $3 \sigma$ likelihood
contours in the $(\Omega_b h^2, \Omega_\phi^{BBN})$ plane.

Our main result is that the experimental data for $^4He$ and $D$ does
not favour the presence of a dark energy component, providing the
strong constraint $\Omega_\phi(\mbox{MeV}) < 0.045$ at $2\sigma$
(corresponding to $\lambda > 9$ for the exponential potential
scenario), strengthening significantly the previous limit
of ~\cite{Ferreira:1998hj}, $\Omega_\phi(\mbox{MeV}) < 0.2$. 
The reason for the difference is due to
the improvement in the measurements of the observed abundances, 
especially for the deuterium, which
now corresponds to approximately $\Delta N_{\mbox{eff}} < 0.2-0.3$
additional effective neutrinos (see, e.g. \cite{burles}), 
whereas Ref.~\cite{Ferreira:1998hj}
used the conservative value $\Delta N_{\mbox{eff}} < 1.5$.

One could worry about the effect of any underestimated systematic
errors, and we therefore multiplied the error-bars of the observed
abundances by a factor of $2$. Even taking this into account, there is
still a strong constraint $\Omega_\phi(\mbox{MeV}) < 0.09$ ($\lambda
>6.5$) at $2\sigma$.

\medskip
{\it Constraints from CMB.} The effects of a scaling field on the angular power spectrum of the
CMB anisotropies are several \cite{AS}. 
Firstly, if the energy density in the scaling
quintessence is significant during the radiation epoch, this would 
change the equality redshift and modify the structure of the peaks
in the CMB spectrum (see e.g. \cite{hu}).
 \begin{figure}[htb]
\begin{center}
\epsfxsize=6.2cm
\epsfysize=6.2cm
\epsffile{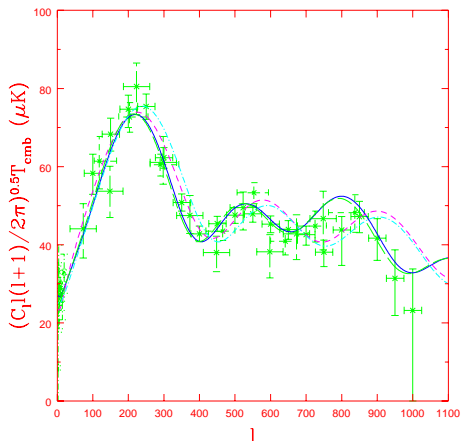}
\caption{CMB anisotropy power spectra for the Albrecht-Skordis models 
with $\lambda=3$, $\phi_0=87.22$ and $A=0.009$ 
(long dash), $\lambda=8$, $\phi_0=32.329$ and $A=0.01$
(short dash) (both with $\Omega_{\phi}=0.65$ and $h=0.65$), 
and a cosmological constant with $\Omega_{\Lambda}=0.65$, 
$N_{\nu}=3.04$ (full line) and $N_{\nu}$=7.8(dash-dot).}
\end{center}
\label{figcmb}
\end{figure}

\begin{figure}[htb]
\begin{center}
\epsfxsize=6.2cm
\epsfysize=6.2cm
\epsffile{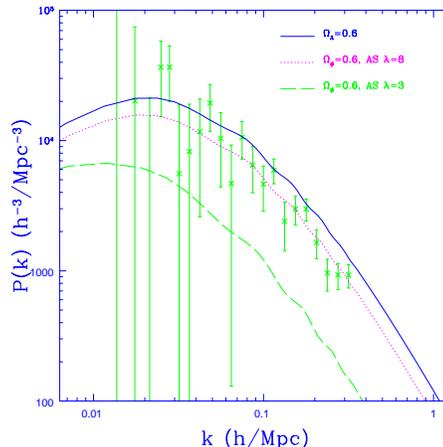}

\caption{Matter power spectra for 3 models in fig. 3.
The predictions support those in the CMB spectra, the quintessence
model in agreement with BBN $\lambda=8$ (short dash) mimics the
$\Omega_\Lambda=0.65$ spectrum with $N_\nu$=3.0 (full line). The model
with $\lambda=3$ (long dash)
is in clear disagreement with observations.}
\end{center}
\label{figmat}
\end{figure}

Secondly, since the inclusion of a scaling field changes the overall
content in matter and energy, the angular diameter distance of the
acoustic horizon size at recombination will change.  This would result
in a shift of the peak positions on the angular spectrum.  It is
important to note that this effect does not qualitatively add any new
features additional to those produced by the presence of a
cosmological constant \cite{eb}.

Third, the time-varying Newtonian potential after decoupling will
produce anisotropies at large angular scales through the Integrated
Sachs-Wolfe (ISW) effect.  Again, this effect will be difficult to
disentangle from the same effect generated by a cosmological constant,
especially in view of the affect of cosmic variance and/or
gravity waves on the large scale anisotropies.

Finally, the perturbations in
the scaling field about the homogeneous solution will also slightly
affect the baryon-photon fluid modifying the structure of the 
spectral peaks. However, this effect is generally negligible.

From these considerations, supported also by recent CMB 
analysis \cite{balbi,bondq}, we can conclude that 
the CMB anisotropies {\it alone} cannot give significant 
constraints on $w_\phi$ at late times.
If, however, $\Omega_{\phi}$ is significant during 
the radiation dominated epoch it would leave a characteristic 
imprint on the CMB spectrum. 
The CMB anisotropies can then provide a useful cross check
on the bounds obtained from BBN.

To obtain an upper bound on $\Omega_\phi$ at last scattering, we perform
a likelihood analysis on the recent Boomerang~\cite{Boom2} and 
DASI~\cite{Dasi} data.
The anisotropy power spectrum from BOOMERang and DASI was estimated
in $19$ bins between $\ell= 75$ and $\ell=
1025$ and in $9$ bins, from $\ell=100 $ to $\ell
=864$ respectively.
Our database of models is sampled as in \cite{esp}, we include
the effect of the beam uncertainties for the BOOMERanG
data and we use the public available covariance matrix and window 
functions for the DASI experiment.
There are naturally degeneracies between $\Omega_m$ and $\Omega_\Lambda$
which are broken by the inclusion of SN1a data~\cite{super1}.

By finding the remaining ``nuisance'' parameters which maximise
the likelihood, we obtain $\Omega_\phi <0.39$ at $2\sigma$ level 
during the radiation dominated epoch.
Therefore, while there is no evidence from the CMB anisotropies
for a presence of a scaling field in the radiation dominated regime, 
the bounds obtained are actually larger than those from BBN.

In Fig.~3 we plot the CMB power spectra for 4 alternative scenarios.
The CMB spectrum for the model which satisfies the BBN constraints is
practically indistinguishable from the spectrum obtained with a
cosmological constant.  Nonetheless, if the dark energy component
during radiation is significant, the change in the redshift of
equality leaves a characteristic imprint in the CMB spectrum, breaking
the geometrical degeneracy. This is also found when considering non-
minimally coupled scalar fields~\cite{RB}, even when the scalar is a
small fraction of the energy density at last scattering.
In the minimally coupled models considered here, this is equivalent to an
increase in the neutrino effective number, i.e. altering the number of
relativistic degrees of freedom at last scattering.

In Fig.~4 we have plotted the corresponding matter power spectra
together with the decorrelated data points of Ref.~\cite{Ham}.  As one
can see, the model with $\lambda=3$ is in disagreement with the data,
producing less power than the model with $\lambda=8$, with this last
one still mimicking a cosmological constant.  The less power can be
still explained by the increment in the radiation energy component
which shift the equality at late time and the position of the
turn-around in the matter spectrum towards larger scales.  A bias
factor could in principle solve the discrepancy between the
$\lambda=3$ model and the data, however, the matter fluctuations over
a sphere of size $8 h^{-1} {\rm Mpc}$ are $\sigma_8 \sim 0.5$ to be
compared with the observed value $\sigma_8=
0.56\Omega_m^{-0.47}\sim0.9$ \cite{viana}.  Weak lensing observations
\cite{vanW} may open up further opportunities to constrain
quintessence models even more tightly through the matter power
spectrum.

\medskip
{\it Conclusions.}  
We have examined BBN abundances and CMB anisotropies in a cosmological
scenario with a scaling field.  We have quantitatively discussed how
large values of the fractional density in the scaling field
$\Omega_{\phi}$ at $T \sim 1 $MeV can be in agreement with the
observed values of $^4He$ and $D$, assuming standard Big Bang
Nucleosynthesis.  The $2\sigma$ limit $\Omega_{\phi}(1 \mbox{MeV}) <
0.05$ severely constrains a wide class of quintessential scenarios,
like those based on an exponential potential.  
For example, for the pure exponential
potential the total energy today is restricted to 
$\Omega_{\phi}={3 \over 4} \Omega_{\phi}(1 \mbox{MeV}) \le0.04$.  
Our $2\sigma$ limit on the $\lambda$ parameter, could also
place useful constraints on other dark energy models.  In the case of
the Albrecht-Skordis model, for example, combining our result with the
condition $A\lambda^2 \le 1$, one finds that $A < 0.02$ in order to
have an eternal acceleration. Furthermore, if we want to have
$\Omega_{\phi} \ge 0.65$, then one must have $\phi_0 < 29$.

As mentioned earlier, our BBN constraint is limited to models assuming
standard Big Bang Nucleosynthesis, where the scaling
field simply adds energy density to the expanding universe.

The bound on $\Omega_\phi(\mbox{MeV})$ can be also weakened  by
introducing new physics which might change the electron neutrino
distribution function. Such distortions will alter the neutron-proton
reactions, and subsequently the final abundances. Such new physics
could take the form of heavy decaying neutrinos, or light electron
neutrinos oscillating with a sterile species, both of which can lead
to fewer effective degrees of freedom (see e.g.
\cite{Dolgov:1999st,DiBari:2001wd} and references therein).

There are several quintessence models which evade the BBN bound, and
let us mention a few. 
The simplest way is to modify the standard model of reheating
in order to have late-entry of the field in the attractor solution,
after BBN \cite{Ferreira:1998hj}.
In the 'tracking oscillating energy' of
Ref.~\cite{Dodelson:2000fq} one can chose parameters which both let
$\Omega_\phi$ be small at BBN times and big today. However, since the
probability of the randomly selected parameters decreases rapidly for
stronger BBN constraints, this model may not appear too natural (in
the language of~\cite{Dodelson:2000fq}). Another class of models which
evade the BBN bound are the models exhibiting kination at early
times. This means that $\Omega_\phi$ is suppressed at early times,
taking a value well below that made by the BBN constraint \cite{BBM}.

All these models are compatible with our $1\sigma$ constraint
obtained from CMB data but they nonetheless 
leave a characteristic imprint on CMB and also large scale structure. 
Inevitably therefore future data from
satellite experiments like MAP or Planck, and measurements of the matter power spectrum using weak lensing and the Digital SLOAN survey will enable even tighter limits to be placed on the presence of a scalar quintessence field
in the early universe.

\medskip
\textit{Acknowledgements} It is a pleasure to thank Ruth Durrer, 
Pedro Ferreira, Gianpiero Mangano, Joe Silk and 
Julien Devriendt for comments and
suggestions. RB and AM are supported by PPARC.  
SHH is supported by a Marie Curie Fellowship of the European Community
under the contract HPMFCT-2000-00607.
We acknowledge the use of CMBFAST~\cite{CMBFAST}.

\end{document}